\documentclass[runningheads]{llncs}
\usepackage{graphicx}
\usepackage{amsfonts}
\begin{document}
\title{Assessment of Left Atrium Motion Deformation Through Full Cardiac Cycle}
%
%
\author{Abdul Qayyum\inst{1,2} \and Moona Mazher\inst{3} \and Angela Lee\inst{1} \and Jose A Solis-Lemus\inst{1}\and \\  Imran Razzak\inst{3} 
 \and Steven A Niederer\inst{1,2} }
\authorrunning{A. Qayyum et al.}
%
\institute{National Heart \& Lung Institute, Faculty of Medicine, Imperial College London, United Kingdom \and
Turing Research and Innovation Cluster: Digital Twins, The Alan Turing Institute, London, United Kingdom \and
Department of Computer Science, University College London, United Kingdom\\
 \and
School of Computer Science and Engineering, University of New South Wales, Sydney, Australia \\
\email{\{a.qayyum,chung.lee,j.solis-lemus,s.niederer\}@imperial.ac.uk} \\
\email{m.mazher@ucl.ac.uk, imran.razzak@unsw.edu.au}
}

\maketitle              
\begin{abstract}

Unlike Right Atrium (RA), Left Atrium (LA) presents distinctive challenges, including much thinner myocardial walls,  complex and irregular morphology, as well as diversity in individual's structure, making off-the-shelf methods designed for the Left Ventricle (LV) may not work in the context of the left atrium. To overcome aforementioned challenges, we are the first to present comprehensive technical workflow designed for 4D registration modeling to automatically analyze LA motion using high-resolution 3D Cine MR images. We integrate segmentation network and 4D registration process to precisely delineate LA segmentation throughout the full cardiac cycle. Additionally, an image 4D registration network is employed to extract LA displacement vector fields (DVFs). Our findings show the potential of proposed end to end framework in providing clinicians with novel regional biomarkers for left atrium motion tracking and deformation, carrying significant clinical implications.
\end{abstract}
\section{Introduction}
The operational and structural integrity of the left atrium (LA) is adversely affected in conditions such as  heart failure (HF) and atrial fibrillation (AF). Currently assessment methods mainly rely on one-dimensional or global changes in wall length of atria, which is estimated from single-plane view of cardiac and quantified as left atrium strain. In contrast to LV, that benefits from various dedicated 3D regional metrics for deformation and motion tracking with known clinical correlates, LA lacks such evaluation and quantification methods \cite{liu2023quantification,mazher2024self}. The potential of 3D full coverage strain measurements for the LA lies in their ability to capture crucial information about tissue bio-mechanics, revealing features such as the presence of stiffness, fibrotic regions in LA and expanding applications in clinics. Similar to LV, these measurements can identify atrial dys-synchrony, offering early detection before functional biomarkers i.e. ejection fraction show abnormalities. Thus, the generation of 3D higher resolution spatio-temporal  map  of LA deformation may give specific  LA pathology signatures, enhancing diagnostic and prognostic performance.  Currently, two main imaging techniques have been used for LA motion tracking: Feature Tracking (FT) \cite{smiseth2022imaging} and 2D Speckle Tracking Echocardiography (STE)\cite{badano2018standardization}. Both FT and STE track the position of regions of interest across the cardiac cycle, utilizing optical flow techniques such as block matching or image registration for alignment . Ultimately, they estimate Displacement Vector Fields (DVFs)\cite{iqbal2024hybrid}, which link material points in myocardium across different phases of cardiac cycle \cite{schuster2016cardiovascular,qayyum2023multiscale}. Mathematical manipulation of the DVFs allows for the estimation of strain tensors. 
Recently, deep learning (DL) based image registration with applications in medical imaging showed promising results for LA motion tracking \cite{fu2020deep,qayyum2024unsupervised,,haskins2020deep,fu2020deep,qayyum2023two}. Hence, various DL based image registration methods have been proposed for analyzing LV deformation across the cardiac cycle from Cine MR images, using architectures such as U-Net \cite{morales2021deepstrain,upendra2020convolutional,reddy2020convolutional}, variational autoencoders \cite{bello2019deep,qin2023generative}, pyramid networks \cite{yu2020motion} or Siamese autoencoders \cite{yu2020foal,qin2018joint}. Typically, displacement fields are learned in an unsupervised manner\cite{iqbal2024hybrid}, employing a spatial resampling module. To enhance performance, segmentation maps has been integrated into cardiac motion tracking networks, either by incorporating segmentation outputs as part of the input or by constraining DVFs to align with target segmentation. Multiple cardiac MRI views (3D and 2D) is also being considered to improve motion estimation. Biomechanics-informed modeling \cite{qin2020biomechanics} has also been used which leverages underlying deformation properties of the heart to regularize DVFs toward more realistic motion \cite{meng2022mulvimotion}.  While research to estimate 3D based DVFs for LV from MR images matured over the years, addressing specific challenges posed by the LA remains a challenging task i.e. LA presents distinctive challenges, including much thinner myocardial walls \cite{varela2017novel}, complex and irregular morphology, and diversity in individual's structure \cite{lopez2023warppinn}, making off-the-shelf techniques designed for the LV unlikely to work in the context of the LA. In this work, we are the first time to present 4D Registration model aided with motion tracking module for full cardiac cycle using three different networks to predict DVF. The key contributions of this work are: \textbf{(i)} propose 3D segmentation model (hybrid CNN and transformer) for full cardiac cycle, trained based on ground-truth ED and ES frames of full cardiac cycle of left atrium dataset, \textbf{(ii)} present 4D registration model for full cardiac cycle using three various networks to predict DVF, \textbf{(iii)} propose motion tracking module and spatiotemporal Mask autoencoder network to utilize the representational capacity of conditional variational autoencoder (CVAE) as a foundation for mapping organ deformations across a low-dimensional space that encapsulates concise representations of respiratory states.

%

\section{Method}
\begin{figure}
    \centering
    \includegraphics[width=10cm, height=8cm]{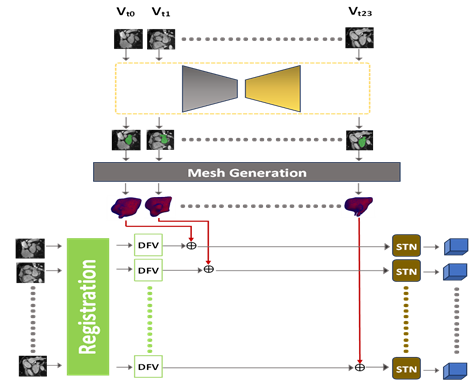}
    \caption{overall pipeline of proposed method for LA segmentation and registration
}
    \label{fig:mesh}
    \vspace{-0.6cm}
\end{figure}

\begin{figure}
    \centering
    \includegraphics[width=12cm]{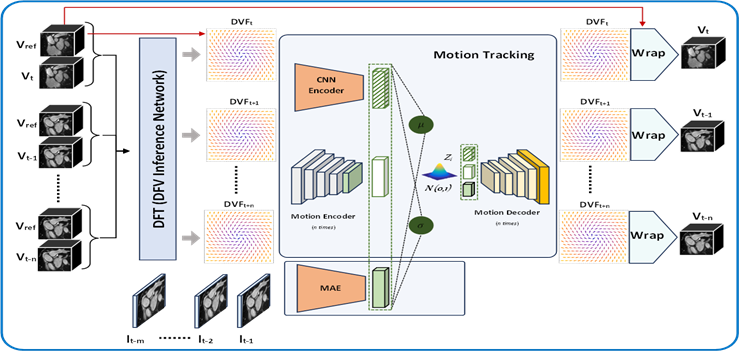}
    \caption{4D Registration method to estimate DVF for LA motion tracking }
    \label{fig:model}
\end{figure}

Figure \ref{fig:model} illustrate the proposed novel framework for estimating left atrial scar or fibrosis throughout the entire cardiac cycle. We utilize 24 frames of 3D short-axis view to segment scar in each 3D short-axis frame using hybrid segmentation (described in sec. \ref{sec:seg}), followed by generating the mesh encompassing the full cardiac cycle based on the segmented frames. We then present 4D registration model (described in section \ref{sec:reg}) and construct DVF for each frame in full cardiac cycle. Finally, DFV and segmented scare of each frame passed to spatial-temporal network to obtain the meshes. Figure \ref{fig:mesh} demonstrate mesh generation through the integration of registration and segmentation. Subsequently, we delve into the model components, offering comprehensive explanations and elucidating their operational intricacies.

\subsection{Fibrosis Segmentation}\label{sec:seg}
We present 3D hybrid fibrosis segmentation framework which consist of transformer aided with attention window-based mechanism in encoder, whereas normal 3D-CNN based module has been used in decoder. We have used rectangular-paralleled-piped windows based swin transformer which is based on computation of attention within the partition of local windows (which are non-overlapping) in low resolution features and original image. It uses paralleled-piped rectangular windows to deal with non-square images through parallel-cross attention. In order to obtain distinctive feature set from each convolutional block in encoder, each block consists of 3D convolutional layers with ReLU activation function and batch-normalization. The three-level deep-supervision is applied to get the aggregated loss between prediction and ground truth. Notice that we have trained model on two frame and used to predict segmentation of full cardiac cycle. 

\subsection{4D Registration Model}\label{sec:reg}
Figure \ref{fig:model} illustrate schematic illustration of our framework, which comprises \textbf{(1)}  motion estimation using DVF inference network, \textbf{(2)} CVAE designed to learn the DVF distribution concerning the surrogate and multi-time predictive module. We formalized modeling task, emphasizing its utility for real-time image-guided motion tracking and estimation for left atrium.

We consider a collection of  $P$ time-resolved 3D acquisitions spanning $k$  respiratory cycles, generating $T$  temporal volumes. Within each subject's dataset $s\in S  V_{t=m}$, we select $V_{t=m}$  at a specific respiratory state $m$  as the reference volume which is utilized to estimate relative-displacements ($\mathbf{dvf_r}$) to other temporal volumes ($V_{t'}$). The motion noticed between fixed $V_t \in R^{H\times W\times D}$ and moving image  $V_m \in R^{H\times W\times D}$, is characterized by a dense displacement field  $dvf_t \in R^{H\times W\times D}$, where $t\in [1,T] \neq m$  and $H$, $W$ and $D$ denotes the height, width, and depth respectively. The objective of motion modelling task is three fold: \textbf{(i)} it involves mapping each deformation to low-dimensional space $dvf_s=dvf_1......dvf_{T-1}$, where $_t \rightarrow z_t \in R^d$ and $d \leq \leq H \times W \times D$ thereby summarizing the observed inputs into compact representation, \textbf{(ii)} it fulfills temporal requirements such that deformations must be forecasted ahead-of-time, \textbf{(iii)} establish a relationship between partial observations and the corresponding extrapolated-in-time dense deformations.

\hspace{-.680cm} \textbf{Motion Tracking Module:} In our scenario, the temporal volumes obtained during free breathing exhibit variations due to tissue deformation caused by complex respiratory motion. The deformations due to complex respiratory motion occur in higher dimensional space, and can be determined by the number of motion and voxels components, whereas spatio-temporal changes are primarily driven by a small set of degrees of freedom, implying that the underlying structure can be represented by a lower-dimensional subspace. Consequently, we can view these variations as points on manifold with significantly fewer dimensions. Besides, Autoencoders offer a solution by leveraging their ability to learn mapping (nonlinear parametric) from volume deformation to latent representations, thereby uncovering these embeddings. 

We have a set of deformations $dvf=\{dvf_1 ....dvf_{T-1}\}$ , and a sequence of 2D slices $I_{seq}=\{I_t,.....I_{t-m}\}$ , and $V_{ref}$ as the reference volume. The objective is to maximize the conditional probability distribution $p_{true} (dvf(dvf|I_{seq})$ to obtain the deformations sequence $dvf\in R^{H \times W \times D \times 3 \times m}$ though, we have partially available information and anatomy of subjects, where $H$, $W$, $D$, and m  represent height, width, depth and the number of predicted time steps, respectively. Accordingly, our goal is to learn parameterized model (with parameters $\theta$) in order to sample new phase-specific deformations, resembling samples from unknown distribution $p$. 

An encoder network is employed to approximate this distribution.
\begin{equation}
    q\phi\big(z_i|I_{seq}, V_{ref}\big) = \mu \big(dvf_i, I_{seq}, V_{ref} \big), \sigma (dvf_i, I_{seq}, V_{ref}\big)
\end{equation}

The network structured with stacked 3D convolutional layers, learns both the mean $\mu \in R^d$  and the diagonal covariance $\sigma \in R^d$  from the data, as described in Figure \ref{fig:model}. During the training, sampling of $z_i$  is made differentiable with respect to $\mu$  and $\sigma$ by employing the reparameterization trick. This is achieved by defining $z_t=\mu+\epsilon * \sigma $  where $\epsilon$ is drawn from a standard normal distribution $\epsilon\sim N(0,1)$.

KL-divergence is employed  in the CVAE, to minimize the distance between the distributions $p_{\theta}$ and  $p_{\phi}$. This KL-divergence loss term is incorporated into the total loss function, which also includes minimizing a reconstruction loss. As part of the spatial warping block, the reference volume undergoes warping (represented by the symbol $o$) using the transformation provided by the decoder. This enables the model to compute a reconstruction term $l_{rec}$ between the warped reference volume ($V_{ref} \quad o \quad dfv_i$) and expected volume $V_t$. We have used the stochastic gradient descent to find the optimal parameters by minimizing the following loss function.

\begin{equation}
    argmin_{\theta} \Bigg[ \frac{1}{n}\sum_{i=1}^n l_{rec} \big(V_{ref}  dfv_t, V_t\big)+KL\big(q_{\phi}(Z_i|I_{seq}, V_{ref}) || P_{\theta (z_i)}\big) \Bigg]
\end{equation} 

\begin{equation}
    l_{rec}=l_{sim}\Big( (V_t, V'_i) + \sigma l_{smooth}(dvf_t) \Big)
\end{equation}

 Where $V'{_t}$ results from warping $V_{ref}$  with estimated motion $dvf$, $\sigma$  is a regularization parameter and $l_{smooth} (dvf_l)\sum_{p\in R^3}|| \delta dvf(p)||^2$ computes the differences between neighbouring 3D position $p$. We applied Fourier band-limited deformation network (Eq.2) to calculate the displacement vector field (DVF) between two frames. Subsequently, we fine-tuned this network to estimate DVF for all frames, effectively serving as an inference network. Notice that, proposed motion tracking framework is designed to be independent of the specific method employed for deformable registration.
 
\hspace{-.680cm} \textbf{Spatio-temporal $2D+t$ Mask autoencoder Network:} We employed a spatio-temporal network that utilizes a 2D+time input sequence which leverages a mask autoencoder pretrained model with an 8-time step configuration to extract features similar to 3D predictive model.

\section{Experiment}
\subsection{Dataset}
We have collected forty-one (41) subjects, each aged thirty-two (32) on average, with an average ejection fraction of 60. The dataset was acquired using a Siemens Aera machine, with voxel sizes ranging from $1.6mm \times 1.6mm$ to $2.1mm \times 2.1mm$ and slice thickness varying between 4mm and 8mm. The images were of dimensions ranging from $210\times 280$ to $400 \times 300$, with the $z$ dimension varying from 50 to 80 slices. For training of segmentation model, manual segmentation of the left atrium was conducted on end-diastolic (ED) and end-systolic (ES) frames. Additionally, manual annotations were available for 5 cases covering the full cardiac cycle, which were utilized to evaluate the segmentation and registration performance.

\subsection{Results and Discussion}
Table 1. shows average Dice coefficients and Hausdorff Distance (mm) on test-set using full cardiac frames. We have compared different variations of proposed framework. We can notice that Motion+MAE encoder achieved higher Dice and lower HD. 

\begin{figure}
    \centering
    \includegraphics[width=8cm]{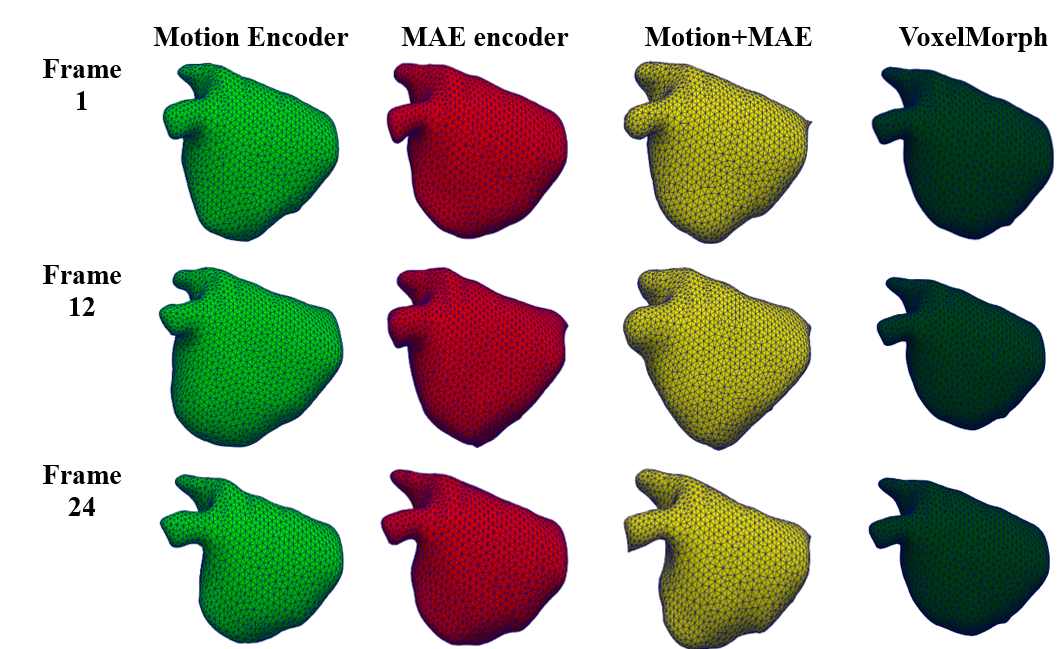}
    \caption{Validation results for subject 1 over the entire cardiac cycle, comparing the performance of the proposed and existing deep learning models.}
    \label{fig:constMesh}
\end{figure}

Figure.\ref{fig:constMesh} shows the visualization of constructed meshes for first frame, mid frame-12 and last frame-24 for subject-1, using proposed Motion Encoder with single encoder (after dvf), proposed MAE encoder and Motion+MAE encoder (after dvf) whereas the last column shows meshes produced after registration using VoxelMorph. To assess the precision of the 4D registration algorithm, we conduct a comparative analysis between the segmented LA following registration and the ground truth LA segmentation for corresponding cardiac phases. We can notice that our model with different variations shows smooth meshes as compared the voxelmorph which showed less surface area.

\begin{figure}
  \centering
  \includegraphics[width=.8\linewidth]{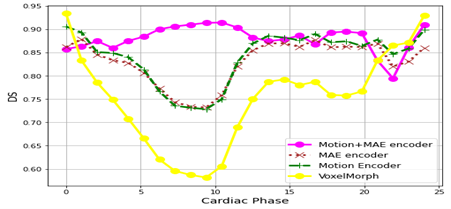}
  \caption{HD for full cardiac cycle using validation subject1 based on proposed and existing registration methods.}
  \label{fig:hd}
\end{figure}

\begin{figure}
  \centering
  \includegraphics[width=.8\linewidth]{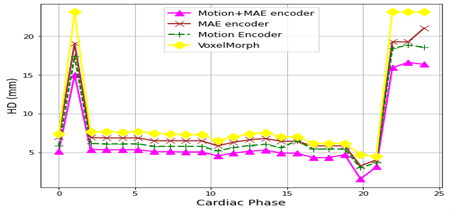}
  \caption{Dice}
  \label{fig:dice}
\caption{ Dice for full cardiac cycle using validation subject1 based on proposed and existing registration methods.}
\label{fig:fig}
\end{figure}

Our primary focus is to utilize DVFs to characterize LA motion throughout the entire cardiac cycle. Figure \ref{fig:constMesh}, \ref{fig:hd} and Figure \ref{fig:dice} shows the comparative analysis of dice and HD score of subject 1 across the entire cardiac cycle, encompassing 24 frames. The visualizations include the ground-truth segmentation post-registration, alongside the moved or predicted segmentation mask (generated after applying Deformation Vector Fields, DVFs). Additionally, a 3D segmentation map of both the ground truth and predicted segmentation for the full cardiac cycle is provided. The proposed 4D registration method consistently outperforms in segmenting each frame, as demonstrated in Figure \ref{fig:constMesh}.

Results showed that by combining the Motion and mask autoencoder (MAE) encoders consistently yielded higher Dice coefficients compared to using individual encoders such as MAE only or Motion only. Conversely, VoxelMorph did not achieve superior performance, as depicted in Figure \ref{fig:hd} and Figure \ref{fig:dice}. Furthermore, the combined Motion and MAE encoder resulted in lower Hausdorff Distance (HD) compared to the individual encoders, as illustrated in Figure \ref{fig:hd}. Integrating the segmentation and image registration tasks, given their interdependence, enhanced the overall performance, as demonstrated for the LV.

Experiment on manual segmentated LA dataset shows that our proposed framework has successfully achieved accurate estimation of high-resolution left atrial (LA) displacement and meshes across the entire cardiac cycle for all 41 subjects. The segmentation model showcases outstanding precision in delineating the LA across diverse phases of the cardiac cycle, displaying superior performance in phases closely aligned with the manually annotated segmentation used for training. Moreover, our 4D registration approach exhibits remarkable accuracy in tracking the LA throughout the cardiac cycle, particularly during phases characterized by substantial displacements.

\begin{table}
\centering
\caption{Average Dice and HD for validation cases using 4D registration Proposed and VoxelMorph methods.}
\begin{tabular}{| l | l | l |}
\hline
Methods & Dice Score & Hausdorff Distance (mm) \\
\hline
\textbf{Motion Encoder} & 0.841± 0.055 & 7.68±4.81 \\
\hline
\textbf{MAE encoder} & 0.829± 0.046 & 8.42± 5.10 \\
\hline
\textbf{Motion+MAE encoder} & 0.880 ± 0.027 & 6.59± 4.27 \\
\hline
\textbf{VoxelMorph} & 0.751±0.100 & 9.57±6.10 \\
\hline

\end{tabular}

\end{table}

Due to our use of a novel high-resolution imaging protocol, our dataset is relatively small, making traditional AI analysis techniques unsuitable. Our model selection addresses the challenges posed by limited data and effectively manages the substantial structural and functional variability of the LA. Training the networks on a subject-by-subject basis mitigates biases from the training data, ensuring continued application in future studies analyzing images from subjects with diverse clinical presentations. Moreover, our proposed method is well-suited for future analyses of the motion tracking and deformation of the right atrium or right ventricle, which is oftenly approximated as thin surfaces.


\begin{figure}
    \centering
    \includegraphics[width=14cm]{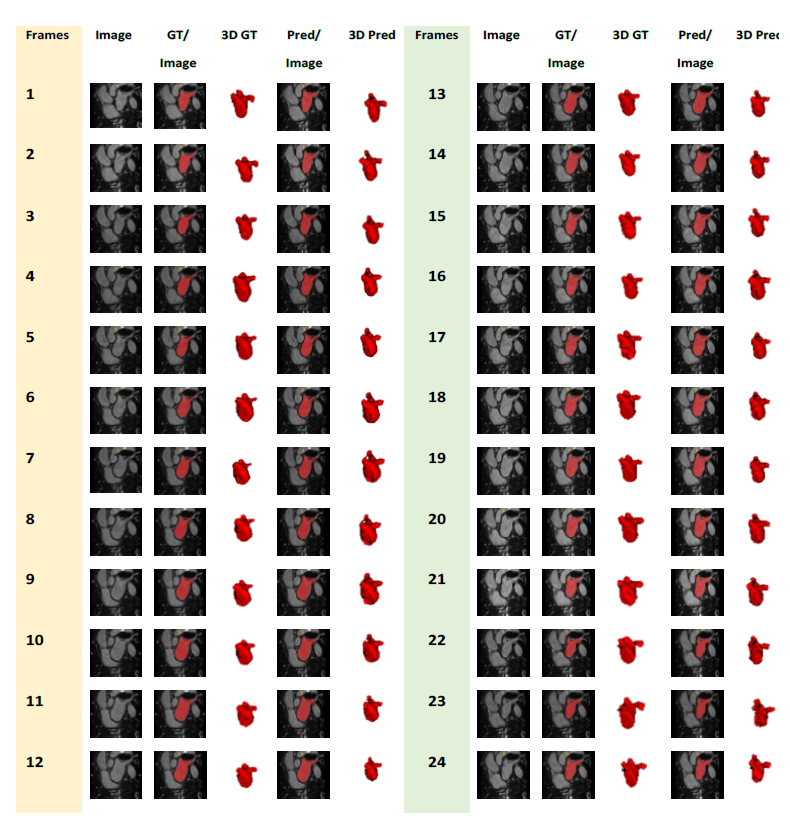}
    \caption{Visualization of results on subject-1 for full cardiac cycle}
    \label{fig:r1}
\end{figure}
Regional analyses of LA motion and deformation hold promise for identifying dense fibrosis or scar tissue in the LA, as these regions typically exhibit greater stiffness than healthy myocardium. Current MRI-based techniques for LA fibrosis detection, relying on late-gadolinium enhancement images, face subjectivity and low reproducibility in analysis. Future investigations will explore whether 3D Cine MRI could serve as an alternative or complementary technique for identifying LA fibrosis, potentially influencing treatment strategies for atrial fibrillation (AF) and the personalization of catheter ablations in AF. Furthermore, leveraging biomechanics-informed regularization, potentially within a physics-informed neural network setting, may offer additional performance improvements. Collectively, these enhancements may contribute to more reliable DVF and strain estimations.


\section{Conclusion}
In this work, we presented a segmentation and 4D registration tool aimed at delivering robust 3D displacement vector fields (DVFs) and strain maps across the entire left atrium (LA), to capture information throughout the full cardiac cycle. An end to end framework demonstrated efficient performance in accurately estimating meshes representing the LA, enabling the derivation of strain and fibrosis metrics from meshes generated after DVFs are applied. Rigorous testing and evaluation using our proprietary full cardiac dataset specific to the LA confirm the effectiveness of our approach. Anticipating that our methodology will facilitate the identification of novel clinical prognostic and diagnostic biomarkers related to LA function, in future, we aim to enhance our understanding of the atria's role in cardiovascular disease processes.

\bibliographystyle{splncs04}
\bibliography{sbibl}

\end{document}